\documentclass[a4paper,11pt]{article}
\usepackage{pos}

\title{Nuclear Potential of Final State Nucleons and Nucleons Plus Pions in  Lepton Nucleus Scattering}

\author*[a]{Arie Bodek}
\author[a]{Tejin Cai}

\affiliation[a]{Department of Physics and Astronomy,  University of Rochester,\\
 Rochester, NY 14627 USA}

\emailAdd{Bodek@pas.rochester.edu}
\emailAdd{tejinc@ur.rochester.edu}

\abstract{
We summarize some of the results presented in   two of our recent papers:  "Removal Energies and Final State Interaction in Lepton Nucleus Scattering"  Eur. Phys. J. C79 (2019) 293 (arXiv:1801.07875[nucl-th],) and "Comparison of optical potential for nucleons and $\Delta$ resonances",  Eur. Phys. J. C80, (2020) 655 (arXiv:2004.00087 [hep-ph]). In addition we address comments made in a paper by U. Mosel, "Comment on "Comparison of optical potential for nucleons and $\Delta$ resonances", e-Print: arXiv:2007.10260 [nucl-th]. 

Within the impulse approximation, the modeling of the energy of final state leptons in  electron and neutrino quasielastic and pion production processes  on nuclear targets in the region of the  $\Delta$ resonance depends on several parameters.  These parameters include the removal energy of the initial state nucleon from the nucleus $\epsilon^{P,N}$, the potentials of electrons,  nucleons and pions in the Coulomb field of the nucleus $|V_{eff}|$, and the kinetic energy dependent nuclear potential for final state nucleons ($U^{QE}_{opt}$)  and for  {\it"nucleon plus pion"} final states in the region of the  $\Delta$ resonance  which we refer to as $U^{\Delta}_{opt}$. We extract these parameters from electron scattering data.  The average removal energies $\epsilon^{P,N}$ are extracted from spectral functions measured in $ee^\prime P$ experiments. $|V_{eff}|$ is extracted from comparisons of electron and positron scattering,  and $U^{QE}_{opt}$ and $U^{\Delta}_{opt}$ are extracted from the peak positions in the energy of final state electrons in QE scattering and pion production in the region of the   $\Delta$(1232) resonance . 

Previous studies have shown that  real part of the optical potential for a nucleon bound in $_{6}^{12}C$ at zero kinetic energy
    $U^{P,N}_{opt}(T=0)\approx$~44~MeV is larger than that for the $\Delta$(1232) resonance $U^{\Delta}_{opt}(T=0)\approx$~30~MeV. We find the reverse at higher kinetic energies. For example at T=100 MeV we find a nucleon potential  $U^{P,N}_{opt}(T=100~MeV)$=20$\pm$5 MeV and  $U^{\Delta}_{opt}(T=100~MeV)$= 30$\pm$5 MeV.  A claim has been made ( arXiv:2007.10260 [nucl-th]) that the our results are inconsistent with the T=0 values.  However, our results are consistent for two reasons. First,  theoretically the kinetic energy dependence of the $\Delta$ potential is flatter than that of the nucleon. Secondly,  in our analysis the extracted $U^{\Delta}_{opt}$ values are the nuclear potential for {\it"nucleon plus pion"}  final states in the region of the  $\Delta$ resonances and therefore includes contributions from both resonance and non-resonance  pion production processes. For Monte Carlo generators that only include the effects of Fermi motion and nuclear potentials,   the relevant parameter is the effective nuclear potential for the  {\it"nucleon plus pion"} final state.

    }
\FullConference{%
  40th International Conference on High Energy physics - ICHEP2020\\
  July 28 - August 6, 2020\\
  Prague, Czech Republic (virtual meeting)
}

\begin{document}
\maketitle

\section{Introduction}

We summarize some of the results presented in   two of our recent papers:  "Removal Energies and Final State Interaction in Lepton Nucleus Scattering"  Eur. Phys. J. C79 (2019) 293 (arXiv:1801.07875[nucl-th],) and "Comparison of optical potential for nucleons and $\Delta$ resonances",  Eur. Phys. J. C80, (2020) 655 (arXiv:2004.00087 [hep-ph]). In addition we address comments made in a paper by U. Mosel, "Comment on "Comparison of optical potential for nucleons and $\Delta$ resonances", e-Print: arXiv:2007.10260 [nucl-th]. 

Within the impulse approximation, the modeling of the energy of final state leptons in  electron and neutrino quasielastic and pion production processes  on nuclear targets in the region of the  $\Delta$ resonance depends on several parameters.  These parameters include the removal energy of the initial state nucleon from the nucleus $\epsilon^{P,N}$, the potentials of electrons,   nucleons and  pions in the Coulomb field of the nucleus $|V_{eff}|$, and the kinetic energy dependent nuclear potential for final state nucleons ($U^{QE}_{opt}$)  and for  {\it"nucleon plus pion"} final states in the region of the  $\Delta$ resonance  which we refer to as $U^{\Delta}_{opt}$. We extract these parameters from electron scattering data.  The average removal energies $\epsilon^{P,N}$ are extracted from spectral functions measured in $ee^\prime P$ experiments. $|V_{eff}|$ is extracted from comparisons of electron and positron scattering,  and $U^{QE}_{opt}$ and $U^{\Delta}_{opt}$ are extracted from the peak positions in the energy of final state electrons in QE scattering and pion production in the region of the   $\Delta$(1232) resonance .  In this paper,  $U^{QE}_{opt}$ refers to  the real part of the nucleon optical potential, and $U^{\Delta}_{opt}$ refers to the  nuclear potential for the  nucleon  plus pion final state  in the region of the  $\Delta$ resonances. Therefore, it includes contributions from both the real part of the optical potential for resonance production and the nuclear potential for-non resonance  pion production processes.
  
    The energy conservation expressions for  QE and  for the production of a {\it"nucleon plus pion"} with  invariant mass W in the region of the $\Delta$ resonance are : 
 \begin{eqnarray}
 \label{Delta_Eq}
  \nu &+&E_i^{P,N} =E_f  \\
  E_i^{P,N}&=& M_{P,N}-\epsilon^{P,N}\\
  E_f^{P} &= &\sqrt{ (\vec k +\vec q_3)^2+M_{P}^2} +U^{QE}_{opt}(T^P)+  |V_{eff}^P| \nonumber\\
  E_f^{N} &= &\sqrt{ (\vec k +\vec q_3)^2+M_{N}^2} +U^{QE}_{opt}(T^N)  \nonumber\\
   T^{P,N} &=& E_f^{P,N} - M_{P,N}\nonumber \\
   E_f^{\Delta+}&=& \sqrt{ (\vec k +\vec q_3)^2+W_{\Delta+}^2} +U^{\Delta}_{opt}(T^{\Delta+}))+  |V_{eff}^{\Delta+}| \nonumber \\
     E_f^{\Delta0}&=& \sqrt{ (\vec k +\vec q_3)^2+W_{\Delta0}^2} +U^{\Delta}_{opt}(T^{\Delta0})\nonumber\\ 
     T^{\Delta(+,0)}&=& E_f^{\Delta(+,0)}-W_{\Delta(+,0)},\nonumber
   \end{eqnarray}
where $\nu$ is the energy transfer to the bound nucleon, $M_P$ is the mass of the proton, $M_N$ is the mass of the neutron,  $W_{\Delta+,0}$  is the  invariant  mass of the  {\it"nucleon plus pion"} final state   in the region of the  $\Delta$ resonance and   $|V_{eff}^{\Delta+}|=|V_{eff}^P|=\frac{Z-1}{Z} |V_{eff}|$.   Here $T^{\Delta(+,0)}$ is the kinetic energy of the resonance of mass $W_{\Delta(+,0)}$ after it leaves the nucleus and is in the same direction as  $\vec{p_{f3}}=\vec k +\vec q_3$, where $ \vec k$ is the momentum of the initial state nucleon and  $\vec q_3$  is the momentum transfer to the bound nucleon.
For a Carbon nucleus we obtain,  $|V_{eff}|$=3.1$\pm$0.25 MeV,   $\epsilon^{P}$ = 27.5$\pm$3 MeV,   $\epsilon^{N}$ = 30.1$\pm$3 MeV.

\section{Examples of fits to data}

\begin{figure*}
  \centering
 \includegraphics[width=2.9in,height=2.6in]
 {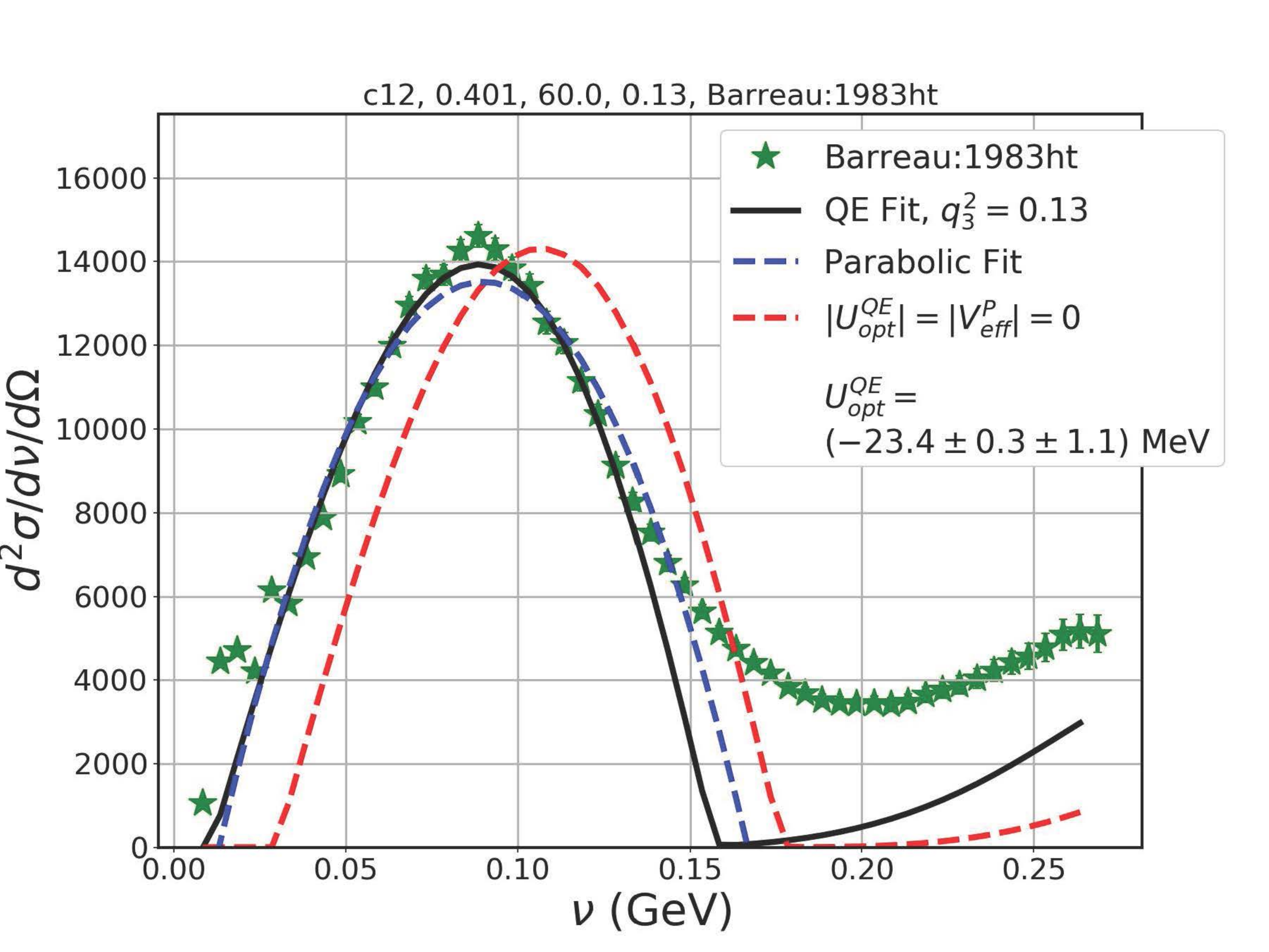}
\includegraphics[width=2.9in,height=2.6in]
{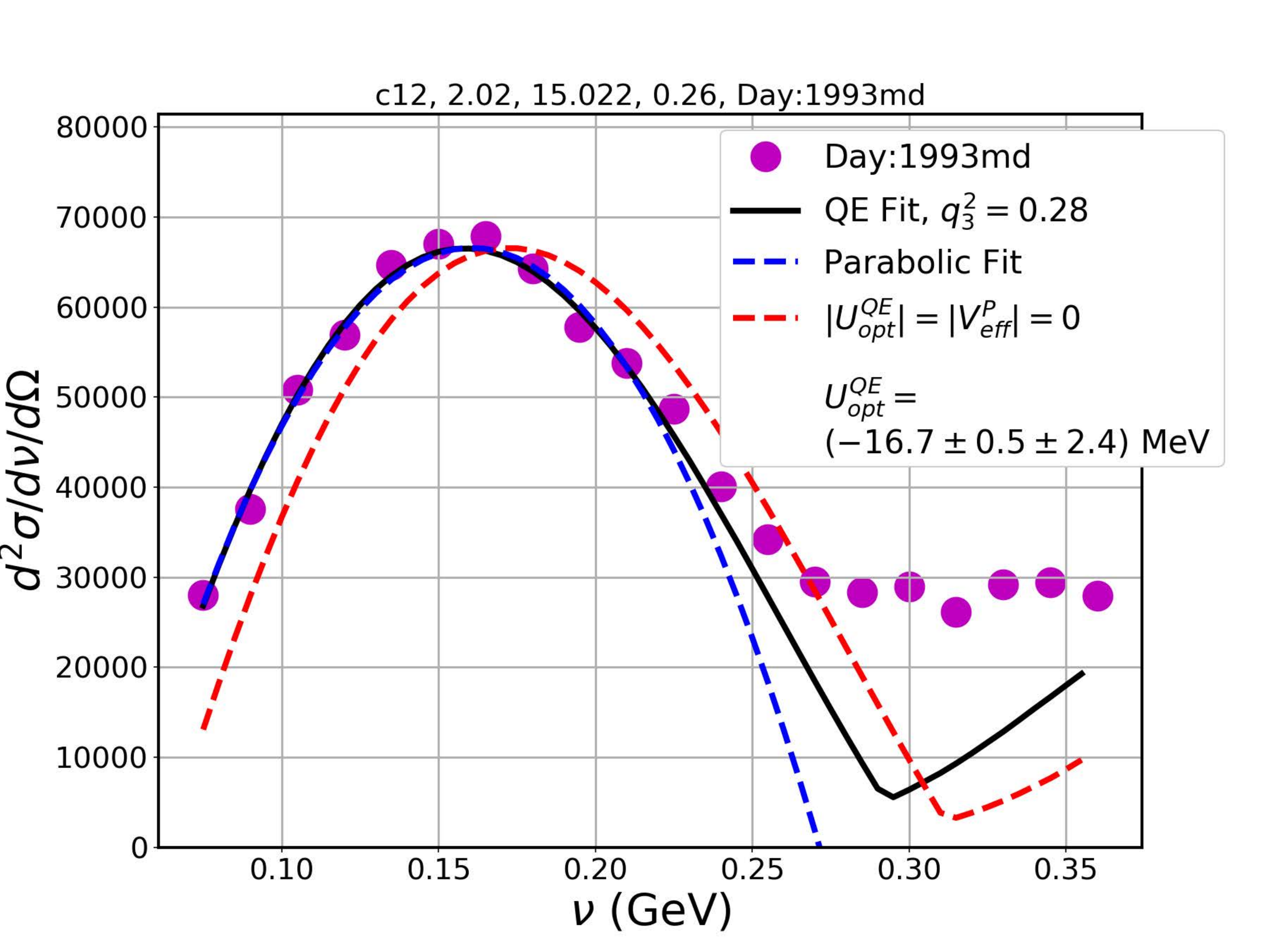}
\caption{
\footnotesize\addtolength{\baselineskip}{-1\baselineskip} 
Examples of fits for  two out of  33  measured   electron scattering QE differential cross sections ob $_{6}^{12}C$. The solid black curves are the RFG fits with the best value of $U^{QE}_{opt}$ for the final state nucleon. The blue dashed curves are simple parabolic fits used to estimate the systematic error.  The difference between $\nu_{peak}^{parabola}$ and $\nu_{peak}^{rfg}$ is used  as a systematic error in our  extraction of $U^{QE}_{opt}$.   The first error shown in the legend is the statistical error in the fit.  The second error is the systematic uncertainty which is much larger. The red dashed curve is the RFG model  with  $U^{QE}_{opt}=0$ and $|V^P_{eff}|=0$.
}
\label{C12_fits}
   \vspace{-0.5cm}
\end{figure*}

Fig. \ref{C12_fits} shows examples of fits for  two out of  33 measured  electron differential cross sections on $_{6}^{12}C$  in the region of the QE peak\cite{paper1}. The solid black curves are the Relativistic Fermi Gas (RFG) fits with the best value of $U^{QE}_{opt}$ for the final state nucleon. The blue dashed curves are simple parabolic fits used to estimate the systematic error.  The difference between $\nu_{peak}^{parabola}$ and $\nu_{peak}^{rfg}$ is used  as a systematic error in our  extraction of $U^{QE}_{opt}$.   The first error shown in the legend is the statistical error in the fit.  The second error is the systematic uncertainty which is much larger. The red dashed curve is the RFG model  with  $U^{QE}_{opt}=0$ and $|V^P_{eff}|=0$.

\begin{figure*}
\centering
 \includegraphics[width=2.9in,height=2.6in]
 {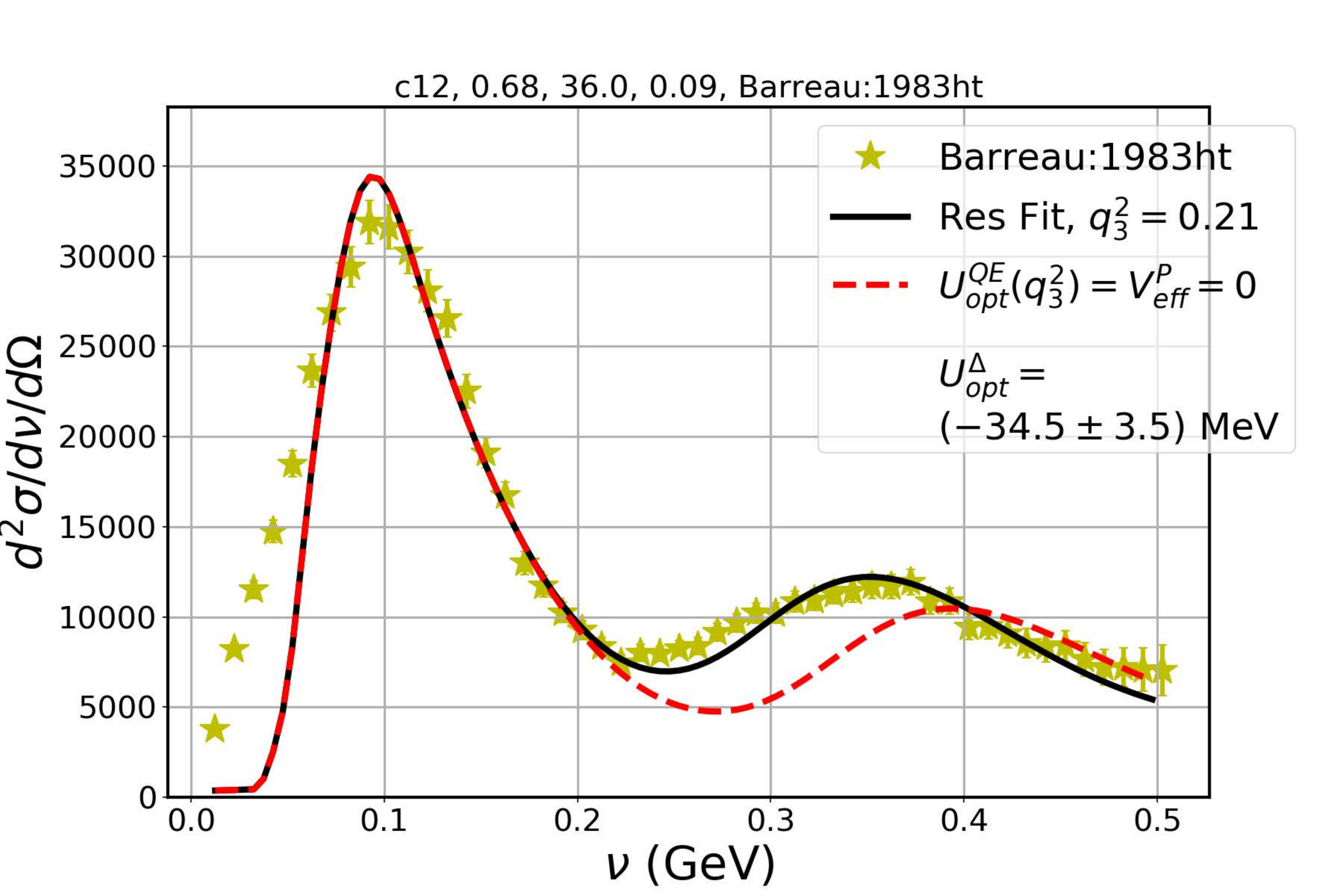}
\includegraphics[width=2.9in,height=2.6in]
{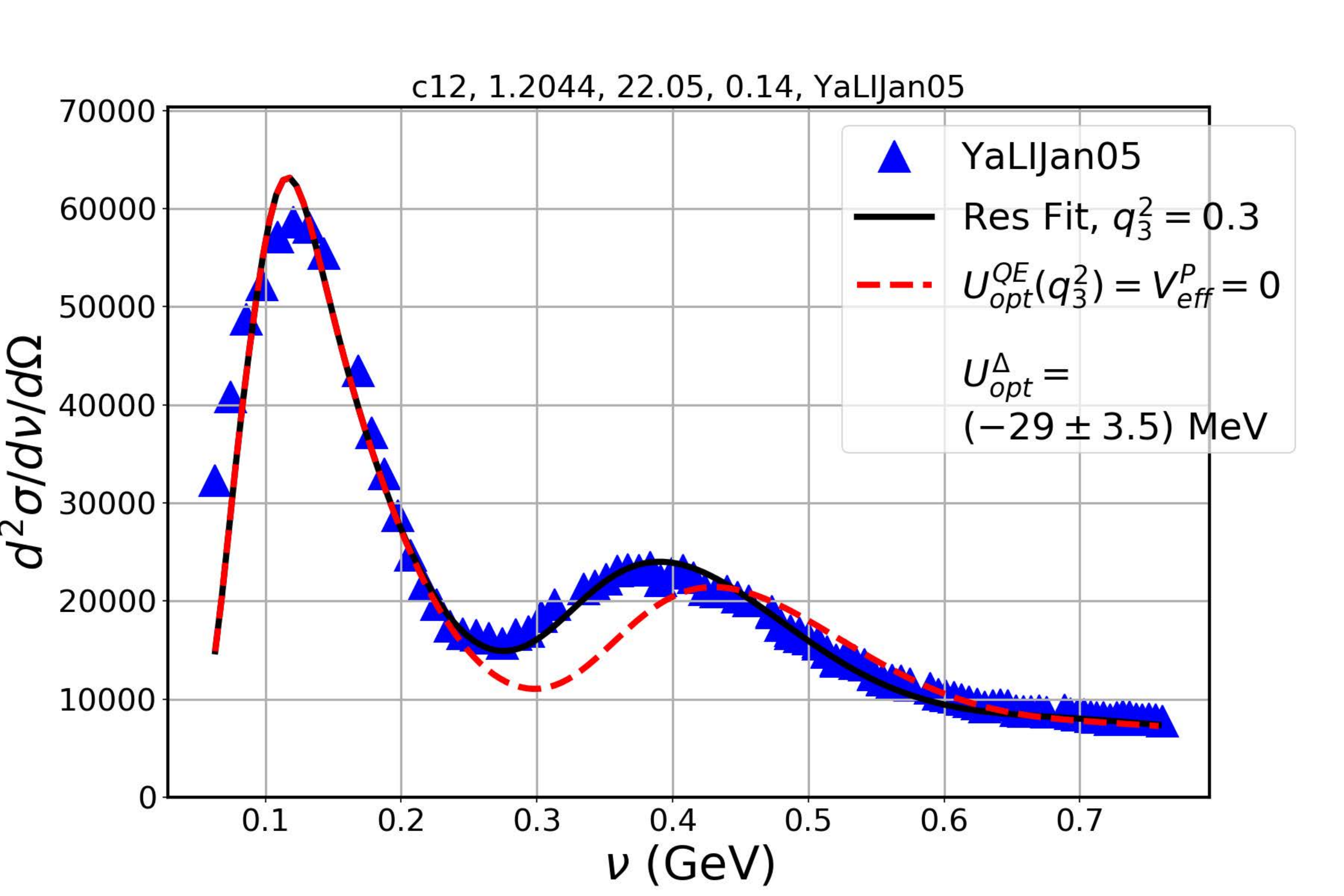}
\caption{
\footnotesize\addtolength{\baselineskip}{-1\baselineskip} 
Examples of fits for  two out of  15  $_{6}^{12}C$  QE and pion production differential cross sections. Here the QE peak is modeled with an effective spectral function (including 2p2h), and pion production is modeled by using RFG to smear fits to resonance and non-resonance pion production structure functions on free nucleons. The solid black curves are the fit with the best value of $U^\Delta_{opt}$. The  dashed red  curves are the predictions  with  $U^\Delta_{opt}=V^\Delta_{eff}=0$.
}
\label{C12_delta_fits}
   \vspace{-0.5cm}
\end{figure*}

Fig. \ref{C12_delta_fits} shows examples of fits for  two out of  15 measured  electron scattering  differential cross sections on $_{6}^{12}C$  for QE scattering and single pion production in the region of the  $\Delta$(1232) resonance\cite{paper2}. Here the QE peak is modeled with an effective spectral function (including 2p2h), and $\Delta$ production is modeled by using RFG to smear fits to resonance and non-resonant pion production structure functions on free nucleons. The solid black curves are the fit with the best value of $U^\Delta_{opt}$. The  dashed red  curves are the predictions  with  $U^\Delta_{opt}=V^\Delta_{eff}=0$. 

   \begin{figure*} 
\centering
  \includegraphics[width=15.cm,height=9.3cm]{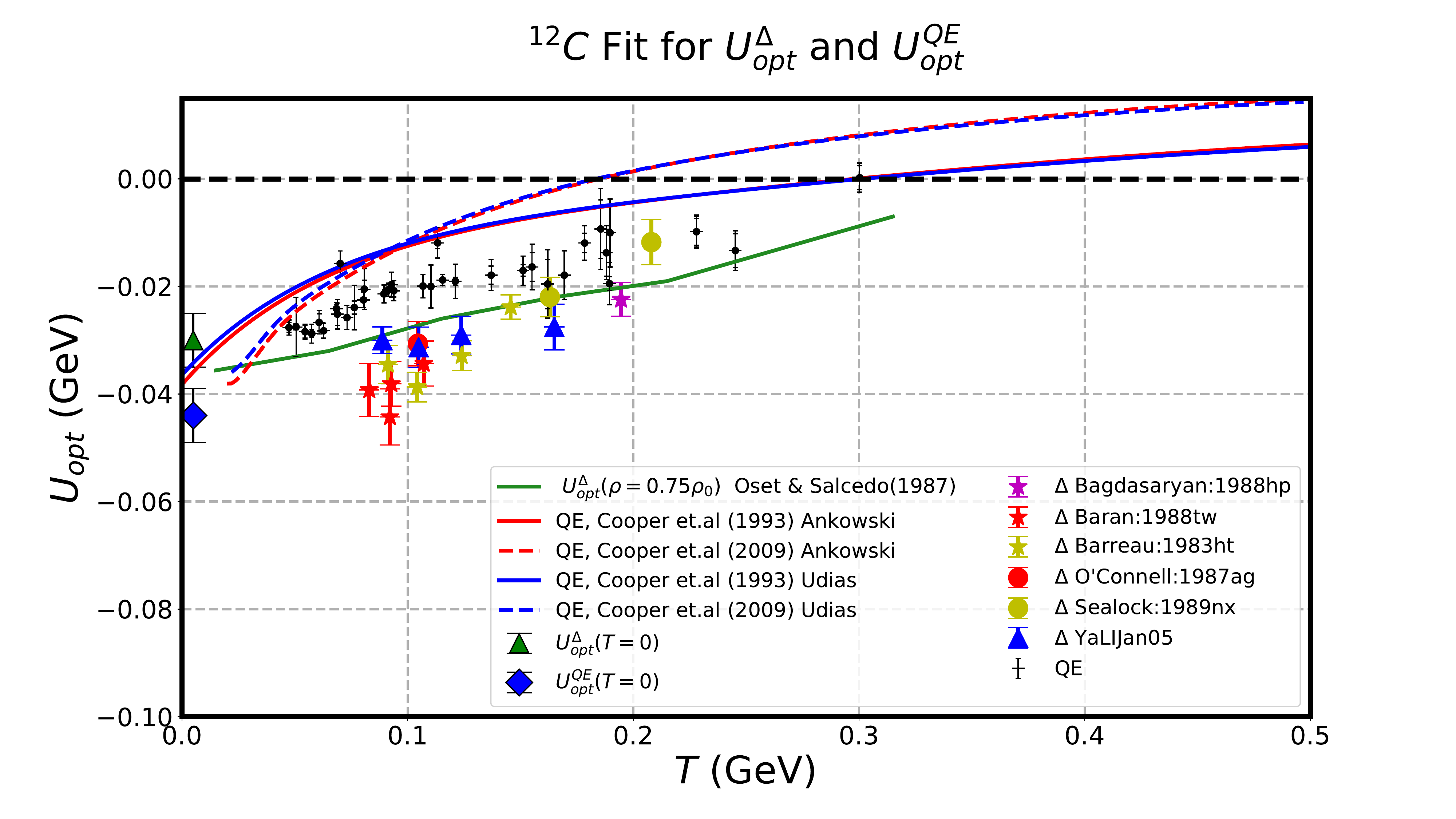}
       \vspace{-0.3cm}
  \caption{
\footnotesize\addtolength{\baselineskip}{-1\baselineskip} 
Extracted values of $U_{opt}^{QE}$  versus kinetic energy  T (small black markers) for  33 $_{6}^{12}C$ and four $_{8}^{16}O$ inclusive electron scattering  spectra.  The curves  are  prediction for $U^{QE}_{opt}$  calculated by Jose Manuel Udias\cite{Udias}(red) and Artur. M. Ankowski\cite{Artur}(Blue)  using the theoretical formalisms of  Cooper 1993\cite{Cooper1993}(solid lines)  and Cooper 2009\cite{Cooper2009} (dashed lines).   
  The larger markers are the  values of the nuclear potential for the  {\it"nucleon plus pion"} final state in the  $\Delta$(1232) region ($U^{\Delta}_{opt}$)  extracted from  15  $_{6}^{12}C$ spectra  for which the measurements extend to higher invariant mass. The solid green curve is the real value of the optical  potential for the $\Delta$ calculated by Oset \& Salcedo (1987) for a nuclear density $\rho =0.75 \rho_0$.   Also shown are   $U^{P,N}_{opt}=U^{QE}_{opt}(T= 0)$=44$\pm$5 MeV,  and  $U^{\Delta}_{opt}(T= 0)$=30$\pm$5 MeV.  }
  \label{potential}
   \vspace{-0.4cm}
\end{figure*}
 \section{Results}
 Fig. \ref{potential} shows the extracted values of nuclear potentials $U_{opt}^{QE}$ and $U_{opt}^{\Delta}$ versus kinetic energy T for  $_{6}^{12}$C.
Shown are values of $U^{QE}_{opt}$ (small black markers) for  33 $_{6}^{12}C$ and four $_{8}^{16}O$ inclusive electron scattering  spectra\cite{paper2}. 
Also shown are  prediction for $U^{QE}_{opt}$  calculated by and Jose Manuel Udias\cite{Udias} and Artur. M. Ankowski\cite{Artur}  using the theoretical formalisms of  Cooper 1993\cite{Cooper1993} and Cooper 2009\cite{Cooper2009}. 
  
 The larger markers are the  values of the nuclear potential for the {\it"nucleon plus pion"} final state in the  $\Delta$(1232) region ($U^{\Delta}_{opt}$)  extracted from  15  $_{6}^{12}C$ spectra  for which the measurements extend to higher invariant mass.
  The solid green curve is the real value of the optical  potential for the $\Delta$  extracted from pion-nucleus reactions by Oset and Salcedo\cite{Oset}. Their curve is  for a nuclear density $\rho =0.75 \rho_0$  where $\rho_0$ is the density of the (interacting) nuclear matter = 0.16 1/fm$^3$.  This value close to the nuclear density for most nuclei including carbon at r=0.  Although this density is larger than the average density for  $_{6}^{12}C$ of 0.62$\rho_0$\cite{Artur} the
 Oset and Salcedo calculation for  $\rho =0.75 \rho_0$  is in  reasonable agreement with the data.  Note that in our analysis
 $U^{\Delta}_{opt}$ refers to the nuclear potential for  the {\it"nucleon plus pion"} final state in the region of the  $\Delta$ resonances and therefore includes contributions from both resonance and non resonance  pion production processes
\subsection {Optical potentials at zero kinetic energy}
 
In some formulations, instead of using a removal energy to describe the initial state, the same optical potential is also used to describe both the initial and final states.
 \begin{eqnarray}
      E_i^{P,N}&=& M_{P,N}-\epsilon^{P,N}\\
     E_i^{P}&=& \sqrt{ (\vec k^2+M_{P}^2}+ U^{P}_{opt} (T^P)+  |V_{eff}^P|   \\
     E_i^{N}&=& \sqrt{ (\vec k^2+M_{N}^2}+ U^{N}_{opt} (T^N) \nonumber
 \end{eqnarray}
For a relativistic Fermi Gas ($K_F$=0.221 GeV/c for carbon)  $T^P=T^N=\frac{K_F^2}{2}$ = 15~MeV at the position of the QE peak ($k_z  \approx 0$). And on average for all $k_z$  $T^P=T^N=\frac{3K_F^2}{5}$ = 12.5~MeV.  Therefore, at small nucleon kinetic energies between 12.5 to 15 MeV we obtain:
 \begin{eqnarray}
 U^{P}_{opt} (T^P=12-15~MeV)& =& -\epsilon^P-T^P-|V_{eff}| = 42.6-45.1 \\
 U^{P}_{opt} (T^N=12-15~MeV) &=& -\epsilon^N-T^N~~~~~~~~~~~~~~= 43.1-45.6 
  \end{eqnarray}
  The above estimate is consistent with nucleon binding potentials of 40-50 MeV which are usually  mentioned in the literature\cite{mosel}.
    $U^{P,N}_{opt}=U^{QE}_{opt}(T^{P,N} )= 0$=44$\pm$5 MeV is shown in In Fig.   \ref{potential}.  Also shown are the T=0 values of 
  $U^{\Delta}_{opt}(T= 0)$=30$\pm$5 MeV\cite{mosel} extracted from an analysis of pion nucleus scattering. 
  
  \section{Conclusion}
  For kinetic energies for which we have measurements on carbon (shown in Fig. \ref{potential})  the following linear functions describe the data:  $U^{QE}_{opt}$(T)=-0.030 +0.092T,   and  $U^{\Delta}_{opt}$(T) =-0.050 +0.161T,  where U and T are in GeV. The results for other nuclei are given in Ref.\cite{paper1} and Ref.\cite{paper2}
  
  The real part of the optical potential for a nucleon at zero kinetic energy
 $U^{P,N}_{opt}(T=0)\approx$44~MeV is larger than the potential for the $\Delta$(1232) resonance $U^{\Delta}_{opt}(T=0)\approx$30~MeV as discussed in Ref. \cite{mosel}.  We find the reverse at higher kinetic energies. For example at T=100 MeV we find a nucleon potential  $U^{P,N}_{opt}(T=100~MeV)$=20$\pm$5 MeV and  $U^{\Delta}_{opt}(T=100~MeV)$= 30$\pm$5 MeV.  
 Unlike the statements  made Ref. \cite{mosel} there is no inconsistency between these two results for two reasons.  First, theoretically,   the kinetic energy dependence of the $\Delta$ potential\cite{Oset} is flatter than that of the nucleon, as seen in Fig, \ref{potential}.
 
Secondly,  in our analysis, the potential is extracted from the location of the QE and $\Delta$ resonance peaks in the differential cross sections, which is of great  importance for the energy reconstruction in neutrino oscillations experiments.  For  the range of kinetic energies in our analysis  the $\Delta$ decays to a nucleon and a pion before it leaves  the nucleus and  resonant amplitude can interfere with the non-resonance single pion production process. For Monte Carlo generators that only include the effects of Fermi motion and nuclear potentials,   the relevant parameter is the effective nuclear potential for the  {\it"nucleon plus pion"} final state.
 
\section{Acknowledgement}
Work supported by the U.S. Department of Energy grant to the University of Rochester (award number DE-SC0008475).

\end{document}